\documentclass[amsmath,amssymb,preprint,aps,pre]{revtex4-2}
\usepackage[utf8]{inputenc}
\usepackage{lipsum} 
\usepackage{graphicx}
\usepackage{dcolumn}
\usepackage{bm}
\usepackage[dvipsnames]{xcolor}
\usepackage[normalem]{ulem} 
\newcommand{\RefereeA}[1]{\textcolor{black}{#1}} 
\newcommand{\vo}[1]{\textcolor{black}{#1}} 
\newcommand{\ef}[1]{\textcolor{black}{#1}} 

\begin{document}

\title{Role of Friction on the Formation of Confined Granular Structures}

\author{Vinícius Pereira da S. Oliveira}
\affiliation{Faculdade de Engenharia Mec\^anica, UNICAMP-Universidade Estadual de Campinas, Rua Mendeleyev, 200 Campinas, SP, Brazil}
\affiliation{Laboratoire PIMM, CNRS, Arts et Métiers Institute of Technology, Cnam, 151 boulevard de l’H\^{o}pital, Paris, France}
\author{Danilo S. Borges}
\affiliation{Faculdade de Engenharia Mec\^anica, UNICAMP-Universidade Estadual de Campinas, Rua Mendeleyev, 200 Campinas, SP, Brazil}
\affiliation{Faculty of Physics, University of Duisburg-Essen, 47057 Duisburg, Germany}
\author{Erick M. Franklin}
\affiliation{Faculdade de Engenharia Mec\^anica, UNICAMP-Universidade Estadual de Campinas, Rua Mendeleyev, 200 Campinas, SP, Brazil}
\author{Jorge Peixinho}
\affiliation{Laboratoire PIMM, CNRS, Arts et M\'etiers Institute of Technology, Cnam, 151 boulevard de l’H\^{o}pital, Paris, France}
\email[Corresponding author: ]{jorge.peixinho@cnrs.fr}

\begin{abstract}

\RefereeA{Unstable systems of fluidized grains in a very-narrow vertical tube} can auto-defluidize after some time, the settling particles forming either a glass- or crystal-like structure. 
We carried out experiments using different polymer spheres, of known friction and roughness, fluidized in water.
A diagram was obtained for the \RefereeA{shell-settled} particles when the coefficient of friction is of the order of 0.1, and their structure is characterized through an analysis of the nearest neighbors' angles.
\RefereeA{We show that the level of velocity fluctuations is higher for the high friction material, and that, once defluidized, the decrease in the coefficient of sliding friction leads to more organized (crystal-like) structures, while those with higher friction coefficients are amorphous (glass-like structures).}
Our results bring new insights for understanding the formation of glass- and crystal-like structures based on the material surface properties.

\end{abstract}

\maketitle

\section{Introduction}

Granular materials are ubiquitous on Earth, Earth's moon, Mars, and other celestial bodies, but their behavior under different conditions is still far from being completely understood, those materials assuming characteristics of either solids, liquids, or gases \cite{Duran,Andreotti_6}. 
Besides acquiring a better knowledge on how granular systems behave, granular materials can be studied for understanding problems at smaller scales, such as gas dynamics \cite{Kohlstedt,Leconte}, interface dynamics \cite{Fan}, and the formation of resisting structures \cite{Cates,Majmudar,Bi}.
In the particular case of transitions from liquid-like (dense regime \cite{Andreotti_6}) to solid-like structures, one problem of interest is how and under which conditions either glasses or crystals appear.
Jammed or glass-like structures of spherical particles were often considered in numerical simulations using frictionless interactions \cite{OHern2003,Mari2014}, where the focus was on the search for critical points and scaling laws.
When taking into account friction, some contact models explicitly demand a friction coefficient for the solid phase, yet simulations are usually performed with an arbitrary friction coefficient.
Hence, there is a lack of data of particulate systems where the friction and roughness are well characterized.

One system that can mimic the glass formation process is a solid-liquid fluidized bed (a suspension of solid particles in an ascending liquid) confined in a tube, as proposed by Goldman and Swinney \cite{Schroter2005,Goldman}. 
They showed that, when defluidizing (decelerating flow), the system can reach a glass structure (although they did not measure the structure in detail) under water velocities still above that for minimum fluidization, which is the minimum cross-sectional average velocity for fluidization. 
In this state, they showed that the structure is roughly static, but the grains have some degree of fluctuation. 
Interestingly, by slightly increasing the water flow from this point, grain fluctuations stop, and the granular structure jams (absence of motion at both, the bed and grain scales). 
In addition, they showed that the formation of this granular glass depends on the deceleration rate of the water (the water flow playing the role of temperature), and proposed that it does not depend on the properties of grains. 
Later, using a similar system but varying the type of grains, C\'u\~nez and Franklin \cite{Cunez3} showed that bed solidification does not depend on the defluidization rate, but on the particle type (diameter, density, and surface roughness), in opposition to Goldman and Swinney \cite{Goldman}. 
In addition, Oliveira et al. \cite{Oliveira} showed that under some conditions, solidification can be intermittent, i.e., solid and fluid-like structures can alternate in the system.

In the following, we investigate experimentally fluidized beds of different particles to understand the effect of solid-solid friction and surface roughness on the glass- or crystal-like structures observed.

\section{Methodology}

\begin{figure}[b]
    \centering
    \includegraphics[width=1.0\linewidth]{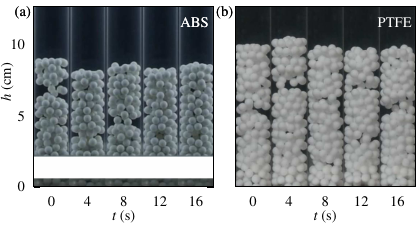}
    \caption{Snapshots of the particle beds made of (a) ABS particles $N=200$ and (b) PTFE particles $N=200$. 
    The white stripe represents an area where data was distorted, hence it is not shown in the figure. The upward water mean velocity is $U=8.7\pm0.1$ cm/s.}
    \label{fig:snapshots}
\end{figure}

\begin{table*}
    \centering
    \begin{tabular}{ccccccc}
    \hline
Sphere material & $d$ (mm) & $\rho$ (g/cm$^3$) & Asphericity (\%) & $R_a$ ($\mu$m) & $\mu_{p-p}$ & $\mu_{p-w}$\\
    \hline
ABS  & $5.91\pm0.01$ & $1.9\pm0.01$ & $0.06\pm0.02$ & $1.25\pm0.60$ & $0.14\pm0.02$ & $0.122\pm0.004$ \\
PTFE & $5.87\pm0.01$ & $2.33\pm0.01$ & $0.11\pm0.04$ & $0.60\pm0.21$ & $0.05\pm0.01$ & $0.078\pm0.012$ \\
    \hline
    \end{tabular}
    \caption{Properties of the spheres: the diameter, $d$, the density, $\rho$, the asphericity, the average roughness, $R_a$, and the measured wetted dynamic sliding friction between the particle and a wall of the same material, $\mu_{p-p}$ and between the particle material and the PMMA wall, $\mu_{p-w}$.}
    \label{tab:tab1}
\end{table*}

Here, our strategy was to monitor how the formation of solid structures in confined solid-liquid fluidized beds depends on the properties of the solid particles, specifically the solid-solid friction coefficient, $\mu$.
For that, we carried out experiments in a water loop, consisting basically of a water reservoir, a centrifugal pump, a flowmeter, a flow straightener, and a tube of diameter $D=25.4$ mm, which has been used in previous works \cite{Cunez3,Oliveira}.
The test section is a vertical transparent tube made of polymethyl methacrylate (PMMA) mounted just downstream the flow straightener (a porous media made of 6 mm particles), as depicted in Fig. \ref{fig:snapshots}.
The fluid was tap water at room temperature, and the resulting flow had a bulk Reynolds number, $Re=\rho_w UD/\eta$, within 1500 and 3300 (typical of transitional pipe flow \cite{Peixinho06}), where $\rho_w$ is the density of water, $U$ is the bulk velocity, and $\eta$ the dynamic water viscosity. The experiments were repeated five times to improve statistics.

The fluidized beds consisted of either spherical particles of acrylonitrile butadiene styrene (ABS) or polytetrafluoroethylene (PTFE). 
The ABS-based particles were used in a previous study \cite{Oliveira}. The particles are confined, with the ratio between the tube diameter, $D$, and that of particles, $d$, within 4.2 and 4.3, and their properties are summarized in Table \ref{tab:tab1}.
The Stokes number based on the terminal velocity $v_t$ of a single particle, $St=v_t d\rho/(9\eta)$, is 467 and 696 for the ABS and PTFE particles, respectively, showing that they have considerable inertia with respect to the fluid and that particle-particle collisions are not negligible \cite{Aguilar}. \ef{The average interstitial velocity, defined as $U/\phi$, varies within 10.6 and 44.2 cm/s, where $\phi=2Nd^3/\left(3HD^2\right)$ is the average packing fraction (global volume concentration of particles), $H$ being the time average of the bed height and $N$ the number of particles in the bed.}
\RefereeA{The incipient velocity is defined as the \ef{bulk} velocity at which particles in the bed begin to move \cite{Cunez3} (oscillations of small amplitude detected between frames)}. 
We observe that above a minimum incipient velocity, $U_{if}\simeq 4.4$ cm/s for PTFE and $U_{if}\simeq 4.5$ cm/s for ABS, the bed fluidizes, and remains indefinitely fluidized for water velocities above a higher threshold, $U_{max}\simeq13$ cm/s. Moreover, either crystal- or glass-like formation can occur when $U_{if} \leq U \leq U_{max}$, and, otherwise, structures in the form of plugs develop with particles migrating from one plug to the other.

\begin{figure}[b]
    \centering
    \includegraphics[width=1.0\linewidth]{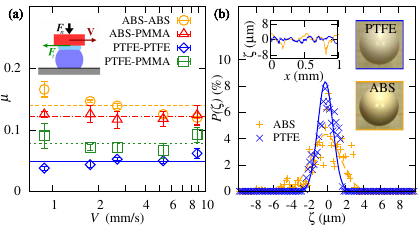}
    \caption{(a) Dynamic coefficient of wet sliding friction for the following particle-plate pairs: ABS-ABS, ABS-PMMA, PTFE-PTFE and PTFE-PMMA. The coefficients are plotted as functions of the sliding velocity $V$. Inset, a schematic sketch of the sliding test. (b)  
    Histograms of roughness $\zeta$ of spherical particles described by a normal distribution. Insets: profilometer traces of $\zeta$ as a function of distance along the surface and images of spherical particles: PTFE and ABS.}
    \label{fig:friction}
\end{figure}

We measured the friction coefficient $\mu=F_t/F_n$ of the materials involved, where $F_t$ and $F_n$ are, respectively, the tangential (measured) and normal (imposed) forces on a sliding sphere.
The dynamic sliding friction measurements were performed using a torsional rheometer Anton Paar MCR 502 with a moving top plate made of PTFE, ABS, or PMMA, and imposing the normal force $F_n$, which can be varied from 0.1 to 5 N. 
A sketch of the setup is shown in the inset of Fig. \ref{fig:friction}(a).
The bottom plate is drilled so that three spheres of PTFE or ABS are flush-mounted. 
The measurements consisted in applying $F_n$ and then measuring the torque $T$ for a constant velocity $V$ of the moving plate.
These parameters are tracked during a given time, corresponding to a total sliding distance of about 9 mm. 
After a short transient, the measurements of $F_n$ and $T$ stabilize.
$F_t$ was computed as the ratio of the torque, $T$, and the distance $R$ from the center of the top plate to that of the contact point with the spheres, $F_t=T/R$, and, afterward, the friction factor $\mu$ was calculated and then averaged for the 20 final time steps.
We used particles of ABS or PTFE and plates of ABS, PTFE or PMMA. 
\RefereeA{The coefficient of friction $\mu$ is slightly dependent on 
$V$ within a range comparable to the collision velocities observed in the fluidized beds. This dependence is stronger for the ABS–ABS pair, where the variation can reach approximately 30\%. The ABS–ABS case appears to exhibit mixed lubrication behavior as described by the Stribeck curve, while the other cases show an almost constant $\mu$, suggesting a regime of soft tribology over this velocity range \cite{Bongaerts2007,Rudge2019}, which is commonly observed in polymers due to the reduced mechanical properties of their free surfaces.}

In addition, the roughness of the particles and plates were measured using a surface profilometer Dektak.
Figure \ref{fig:friction}(b) presents the roughness dispersion of both particles. 
It is represented as a normal-distribution fit of the measured roughness height, $\zeta$.
The surface of PTFE particles exhibits a narrower distribution of $\zeta$ than that of ABS, indicating that PTFE particles are smoother.

The system is analyzed in both the bed and grain scales by processing sequences of images with resolutions of approximately 100 px $\times$ 960 px, and acquired at 30 Hz for 300 s (more details are available in the Supplemental Material \cite{supp} and in Ref. \cite{Oliveira}). The field of view was 25.4 $\times$ 244 mm, so that the spatial resolution was approximately 3.9 px/mm.
In addition, the positions of the particles are tracked and a Voronoi tesselation analysis was carried out for the nearest neighbors' angles, in order to identify the two-dimensional circumferential packing structures along the wall. 
A narrow angle distribution is a signature of a crystal-like structure.
\RefereeA{No calibration for correcting optical distortions due to the tube wall was performed\ef{, and to minimize this effect} particles located at a distance greater than one particle diameter $d$ from the \ef{wall were} neglected in the Voronoi tesselation. \ef{We note that unwrapping the acquired images introduced small uncertainties that deteriorated the data, so that we decided not to do it.}}

\section{Results}
The results consist of friction and roughness measurements of the particles, tracking of the particle motion at both large scale and small fluctuations, a phase diagram for the formation of static glass- or crystal-like structures, and the associated timescales. Finally, the static structures are analyzed through angles of contact chains. \ef{We note that a dataset containing experimental data and numerical scripts for processing the data is available on an open repository \cite{Supplemental2}.}

\subsection{Friction and roughness}

Figure \ref{fig:friction}(a) presents the friction coefficient $\mu$ as a function of the velocity $V$ of the moving plate for the following couples of wetted spheres and plates, respectively (a drop of water was added between the sphere and the plate): PTFE-PTFE, PTFE-PMMA, ABS-PMMA, and ABS-ABS.
The range of used velocities is smaller than in the fluidized bed, lower velocity of the plate inducing stick-slip instabilities, whereas larger velocities lead to larger fluctuations for the range of normal forces considered.
Overall, a constant friction $\mu$ is measured and the friction level depends on the particle-plate couple of materials.
The lowest value measured was $\mu_{\rm PTFE-PTFE}=0.05\pm0.01$ and the highest $\mu_{\rm ABS-ABS}=0.14\pm0.02$.
The relative variation of the friction is merely a factor 3 and can be explained by the roughness of the spheres, which is described by the arithmetic average of roughness height $R_a=\frac{1}{l}\int^{l}_{0}\left|\zeta\right|dx$, where $l$ is the sliding distance of the probe.
Table \ref{tab:tab1} shows also the asphericity and the coefficient of wetted friction between the particle and the wall, the friction measurements being in agreement with previous works \cite{Farrell2010,Farain2022,Farain2023,Terwisscha-Dekker2023}. 
We note that the coefficients for wetted friction are within the error bars of those for dry friction (see Table S1 of the Supplemental Material).
For PTFE particles in PMMA tube, $\mu_{\rm PTFE-PMMA}>\mu_{\rm PTFE-PTFE}$, so it suggests the PTFE particles tend to crystallize on the wall more easily.

\subsection{Bed motion}
\begin{figure}
     \includegraphics[width=1.0\linewidth]{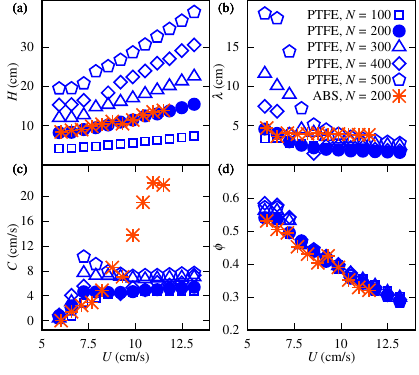}
     \caption{Time averages of particle beds for (a) bed height, $H$, (b) plug length, $\lambda$, (c) celerity, $C$ and (d) global packing concentration, $\phi$, as functions of the upwards water bulk velocity, $U$.}
     \label{fig:dynamics}
\end{figure}

\vo{
In beds with $D/d \leq$ 10, it has been reported that confinement leads to the formation of alternating regions with \ef{high and low particle fractions, known as plugs and bubbles, respectively \cite{Cunez,Cunez3}. The presence of arches within plugs has been reported in previous numerical works \cite{Cunez, Cunez2}; however, it is not possible to directly identify such arches from our front-view images.}
\ef{The plug-bubble structures propagate upward with a typical celerity (velocity), computed as the difference of the plug position (center of mass) between consecutive frames, divided by the corresponding time interval. 
}}
Figures \ref{fig:dynamics}(a-d) show, respectively, the time averages of bed height, $H$, plug length, $\lambda$, plug celerity, $C$, and global volume concentration \ef{$\phi$}, as they converge over time. 
The volume concentration represents the ratio between the volume of the particles and that occupied by the bed as a function of the bulk velocity of the water, $U$, parametrized by the number of particles $N$.
$N$ varies from 100 to 500 for PTFE, and these beds are compared with that of ABS with $N=200$, extracted from Oliveira et al. \cite{Oliveira}.
For all beds tested, we observe that their height $H$ increases while the plug length $\lambda$ decreases with increasing $U$, $\lambda$ reaching a plateau of approximately 1.6 to 3.6 cm.
\RefereeA{The increase of $H$ with $N$ is expected since the size of the system increases (more grains \ef{imply} a higher bed) and, as the bed expands, the concentration decreases (reflected in the packing fraction in Fig. \ref{fig:dynamics}(d)).} 
However, $\lambda$ presents a more complex and non-linear behavior, depending on $U$ and $N$ (see also Fig. S1(b)). \ef{We observe in the experiments that the bed is fluidized in an approximately homogeneous regime when water velocities are just above that for incipient motion, and, by increasing the water velocity, the bed forms smaller plugs. From a certain water velocity on, the size of plugs remains roughly stable, as we can see in Fig. \ref{fig:dynamics}(b).}
By comparing the behaviors of PTFE and ABS particles as a function of $U$ (when $N=200$), we notice that $H$ follows the same trend and presents the same values, but for $\lambda$ the ABS beds have smoother variations and slightly higher values. 
The lower values of $\lambda$ with PTFE particles are probably due to their lower coefficient of friction. 
For the plug celerity $C$, data for PTFE particles show a tendency to reach a plateau, while \ef{data do not allow us to conclude the same for the ABS particles (they keep increasing with $U$, and a plateau might possibly appear at high values of $C$). 
} 

\subsection{Phase diagram}

We observed that, depending on the water velocity $U$, and the number $N$ and type of particles, the bed was either fluidized, formed a static structure (that could be more or less organized, and that we call crystal- and glass-like, respectively), or was in a transient regime \ef{that was metastable,} where \ef{the bed alternatively forms a static structure and refluidizes} spontaneously \cite{Oliveira} (see Supplemental Material Video S1 for an example of the formation of the crystal-like structure \ef{and Video S2 for an example of the metastable regime}). As the crystallization/glass transition takes place, it is noticed that the particles form a circumferential layer along the wall, in part because of the lubrication forces close to the wall \cite{takeuchi2019}, whereas the core of the flow remains fluidized for a certain duration.



As we will see in the following, depending on the friction coefficient of particles, the circumferential structure can be more or less organized, reaching a crystal- or glass-like structure. 
Similar structures have been observed in previous works investigating closed-packed particles confined in tubes with $D/d\leq4$ \cite{Fu2016}.
In the following, we qualify those structures by making use of Voronoi tessellation and the nearest neighbors' angles. 

\begin{figure}
   \centering
   \includegraphics[width=0.7\linewidth]{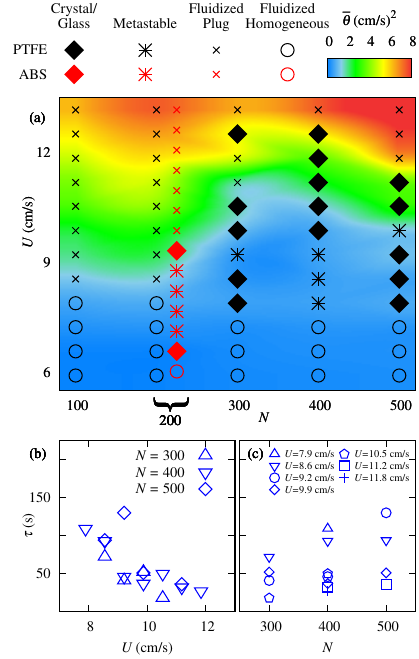}
   \caption{\vo{(a) \ef{Regime map of the time average of ensemble-granular temperature $\overline{\theta}$ in the $N-U$ space, showing the fluidized,} crystal/glass, and metastable structures, for the PTFE particles (the regimes for the ABS particles are shown in red symbols). (b) Duration of crystal formation, $\tau$, as a function of $U$, for different values of $N$. (c) $\tau$ as a function of $N$, parameterized by $U$.}}
   \label{fig:crys-time}
\end{figure}

\ef{Figure \ref{fig:crys-time}(a) shows a regime map of time-averaged granular temperatures computed for the ensemble of particles in the $N-U$ space.} The ensemble-average granular temperature \cite{Andreotti_6, Cunez4} is defined by Eq. (\ref{temp_def}): 

\begin{equation}
    \theta(t) = \frac{1}{2M}\sum_{i=1}^{M}\left({u^\prime}^2_{i}+{v^\prime}^2_{i}\right), 
    \label{temp_def}
\end{equation}

\noindent where $u^\prime_i=u_i-\overline{u}$ and $v^\prime_i=v_i-\overline{v}$ are, respectively, the horizontal and vertical velocity fluctuations of the $i^{\rm th}$ particle, $u_i$ and $v_i$ being the velocities of each particle $i$ and $\overline{u}$ and $\overline{u}$ the corresponding spatial averages\ef{, and} $M$ is the number of analyzed particles. \ef{The time average of the ensemble-averaged granular temperature, $\overline{\theta}$,  is then computed by Eq. \ref{temp_time_avg},}

\begin{equation}
    \ef{\overline{\theta} = \frac{1}{T}\sum_{t=0}^{T}\left(\theta \right),}
    \label{temp_time_avg}
\end{equation}

\noindent \ef{where $T$ is the total duration of the experiment.}

\begin{figure}
    \centering
    \includegraphics[width=0.7\linewidth]{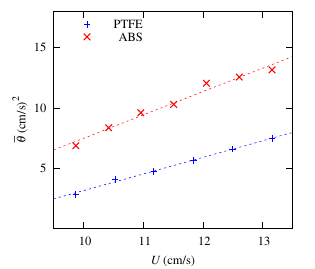}
    \caption{Global granular temperature as a function of the bulk velocity, $U$, in the fluidization regime. The blue points represent PTFE experiments, $N=$ 200, and the red crosses represent the ABS bed with $N=200$. The dashed lines are linear fits. For ABS, the slope is 1.93 cm/s and for ABS, the slope is 1.37 cm/s.}
    \label{fig:temp}
\end{figure}

\ef{In this map, we distinguish the crystal and glass, fluidized, and metastable regimes, with the fluidized regime presenting two sub-regimes: a homogeneous and a plug fluidization. In the homogeneous sub-regime the bed expands without forming granular plugs (sometimes the bed is only broken into two large portions), and it occurs when water velocities are just above that for incipient motion. The formation of plugs, and then the plug sub-regime, takes place for higher water velocities. In general,} for a number of particles $N<300$, the fluidized regime is dominant.
As the number of particles increases, from $N=200$ for ABS or $N=300$ for PTFE particles, crystal/glass and metastable states are observed in a range of $U$. 
The metastable state forms crystal- or glass-like states that are transient \ef{(they form and break spontaneously along the duration of the test)}, whereas the crystal- and glass-like structures are static and steady. \ef{We note that each experiment lasts for 300 s, so that in the crystal- or glass-like states the bed defluidizes before the end of the 300 s and remains defluidized until the end of the experiment. In the metastable state the bed spontaneously alternates between the defluidized and fluidized states along the 300 s.}
In the regime map, metastable states were observed to coexist with crystal- and glass-like structures at some intermediate $U$.

We measured the time interval, $\tau$, from the start of crystal- or glass-like formation until when the structure reaches its final stable state, and it is shown in Fig. \ref{fig:crys-time}(b) as a function of $U$, from which we can observe that
$\tau$ decreases with $U$. 
Figure \ref{fig:crys-time}(c) shows $\tau$ as a function of $N$ for different values of $U$, and we observe that, as expected, the duration increases with the size of the bed. 

In the fluidized regime, the granular temperature increases linearly with $U$ (see Fig. \ref{fig:temp}), in the range of tested velocities, the average temperatures being higher for the ABS beds. 
Therefore, the ABS beds experience higher temperature gradients during defluidization in comparison with the PTFE beds, leading to the formation of glass-like structures.

\subsection{Vonoroi tesselation}

\begin{figure}
     \includegraphics[width=0.8\linewidth]{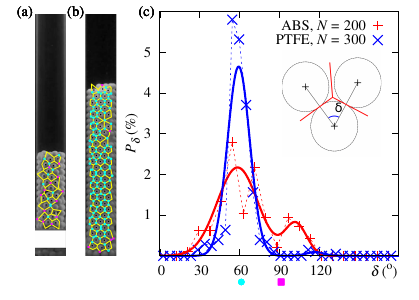}
     \caption{(a) Snapshot of a metastable bed and Voronoi tesselation in yellow of ABS $N=200$ and $U=8$ cm/s and (b) snapshot of a crystallized bed and Voronoi tesselation of PTFE $N=300$ and $U=8$ cm/s.
     Cyan points represent the vertices associated with neighbors' angles of $60\pm5$\textdegree and magenta square points represent the vertices associated with neighbors' angles of $90\pm10$\textdegree.
     (c) Histogram of distribution of nearest neighbors' angles $\delta$ for ABS and PTFE  crystallized packings.
     The continuous lines are double Gaussian fits. The insert is a schematic of the nearest neighbors' angles $\delta$.}
     \label{fig5}
 \end{figure}

We now assess the crystal and glass structures using Voronoi tessellation \cite{Reis2006}, for which we consider only particle centers that are not at the border of images. 
We note that the images are two-dimensional, so that some distortions are expected (although assumed not affecting our results). 
Figures \ref{fig5}(a-b) show examples of Voronoi tesselation (yellow lines) computed for the ABS and PTFE beds, respectively. 
From that, we can observe a less regular and glass-like structure for the ABS, while the structure is more regular for the PTFE, for this reason we call the latter crystal and the former glass in this work. 
From the Voronoi cells, the centers and the angles between the nearest neighbors', $\delta$, can be calculated, and the distribution of $\delta$ is displayed in Fig. \ref{fig5}(c).
For crystallized PTFE states, a single peak distribution is obtained and is fitted with the most probable $\delta=59\pm7$\textdegree. 
For ABS glass states, two peaks are present and fitted with a double Gaussian model, with $\delta=62\pm16$\textdegree and $108\pm15$\textdegree.
For both distributions, the main peak is around $\delta \approx 60$\textdegree, which is typical of a hexagonal packing, and the second for ABS is $\delta \approx 90$\textdegree, which is perhaps related to square packing. 

\begin{figure}
    \includegraphics[width=0.8\textwidth]{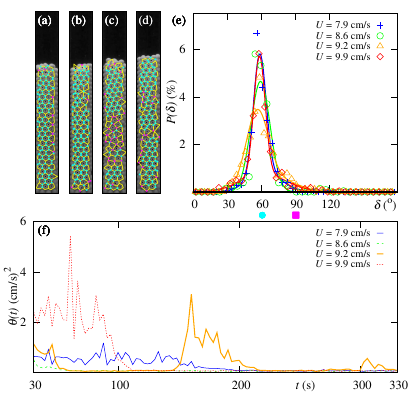}
    \caption{Snapshots of the crystallized bed of PTFE $N=300$ particles experiments and the associated Voronoi diagrams: (a) $U=7.9$ cm/s, (b) $U=8.6$ cm/s, (c) $U=9.2$ cm/s and (d) $U=9.9$ cm/s. 
    Cyan points represent the vertices associated with neighbors' angles of $60\pm5$\textdegree and magenta square points represent the vertices associated with neighbors' angles of $90\pm10$\textdegree.
    (e) Histograms of nearest neighbors' angles for different $U$. 
    The solid lines represent the fits of the histograms. \ef{(f) Ensemble-averaged granular temperatures $\theta$ as a function of time $t$ for the same cases shown in panels (a)-(e).}}
    \label{fig7}
\end{figure}

For PTFE, Figs. \ref{fig7}(a-d) present snapshots of the crystal-like structures for $7.9\leq U \leq 9.9$ cm/s.
The tessellation (also represented as yellow lines) and the nearest neighbors' angles (represented by colored points) of the structure exhibit hexagonal packing.
When varying $U$, one can notice that the hexagonal crystallized structure in Fig. \ref{fig7}(c) is discontinuous from bottom to top as indicated by the magenta points (representing $\delta=90\pm10$\textdegree) and it is interesting to note that this discontinuous configuration is associated to a metastable state (see Fig. \ref{fig:crys-time}(a)).
This evidences the fragility of the structure associated to metastability\ef{, not seen in Figs. \ref{fig7}(a), \ref{fig7}(b) and \ref{fig7}(d), which form persisting crystal-like structures. In all those cases,} the tendency to have a single large peak (see Fig. \ref{fig7}(e)) indicates the formation of hexagonal structures, which are independent of $U$. \ef{Finally, Fig. \ref{fig7}(f) shows the time evolution of the ensemble-averaged granular temperatures $\theta$ of these same four cases, from which we observe that for $U$ $=$ 7.9, 8.6 and 9.9 cm/s $\theta$ decreases and remains within values close to zero, while for the metastable case ($U$ $=$ 9.2 cm/s) $\theta$ alternates between high and low values.}

\section{Conclusions}
In conclusion, fluidization of very-narrow beds consisting of regular particles can lead to fluidization, crystallization/glass formation, or metastable (transient fluidization) regimes, depending on the fluid velocity and number of particles. 
Within a given range of fluid velocities, the bed is in metastable or crystal/glass regimes, the metastable regime becoming more rare as the fluid velocity is increased, with the crystal/glass formation time decreasing also with $U$. Above this range of $U$, the bed is fluidized. The decrease in the coefficient of sliding friction of particles leads to more organized structures, which we associate with crystals, while those with higher friction coefficients are amorphous, which we associate with glasses. 
In addition, we show that the (granular) temperature is an important parameter, with high rates of decrease giving origin to amorphous structures, just as it happens for glasses and crystals.
Besides fluidization and defluidization, our results represents new insights into the formation of crystals and glasses.

\begin{acknowledgments}

We acknowledge the support of FAPESP-CNRS Grant n$^{\rm o}$ 2024-02440-2. EMF, VPSO and DSB are grateful for the support of FAPESP (Grant Nos. 2018/14981-7, 2020/00221-0, 2022/01758-3, 2024/02440-2, 2024/13295-3).

\end{acknowledgments}

\bibliography{bibi}

\newpage

\begin{center}
    Supplemental Material: \\ ``Role of Friction on the Formation of Confined Granular Structures"
    \\
    \vspace{0.5cm}
    Vinícius Pereira da S. Oliveira\\
    \it{Faculdade de Engenharia Mec\^anica, UNICAMP-Universidade Estadual de Campinas, Rua Mendeleyev, 200 Campinas, SP, Brazil}
    \it{Laboratoire PIMM, CNRS, Arts et Métiers Institute of Technology, Cnam, 151 boulevard de l’H\^{o}pital, Paris, France} \\
    \vspace{0.5cm}
    Danilo S. Borges\\
    \it{Faculdade de Engenharia Mec\^anica, UNICAMP-Universidade Estadual de Campinas, Rua Mendeleyev, 200 Campinas, SP, Brazil}
    \it{Faculty of Physics, University of Duisburg-Essen, 47057 Duisburg, Germany} \\
    \vspace{0.5cm}
    Erick M. Franklin\\
    \it{Faculdade de Engenharia Mec\^anica, UNICAMP-Universidade Estadual de Campinas, Rua Mendeleyev, 200 Campinas, SP, Brazil} \\
    \vspace{0.5cm}
    Jorge Peixinho\\
    \it{Laboratoire PIMM, CNRS, Arts et M\'etiers Institute of Technology, Cnam, 151 boulevard de l’H\^{o}pital, Paris, France}
\end{center}

\maketitle

\title{Supplemental Material: \\ ``Role of Friction on the Formation of Confined Granular Structures"}

\author{Vinícius Pereira da S. Oliveira}
\affiliation{Faculdade de Engenharia Mec\^anica, UNICAMP-Universidade Estadual de Campinas, Rua Mendeleyev, 200 Campinas, SP, Brazil}
\affiliation{Laboratoire PIMM, CNRS, Arts et Métiers Institute of Technology, Cnam, 151 boulevard de l’H\^{o}pital, Paris, France}

\author{Danilo S. Borges}
\affiliation{Faculdade de Engenharia Mec\^anica, UNICAMP-Universidade Estadual de Campinas, Rua Mendeleyev, 200 Campinas, SP, Brazil}
\affiliation{Faculty of Physics, University of Duisburg-Essen, 47057 Duisburg, Germany}

\author{Erick M. Franklin}
\affiliation{Faculdade de Engenharia Mec\^anica, UNICAMP-Universidade Estadual de Campinas, Rua Mendeleyev, 200 Campinas, SP, Brazil}

\author{Jorge Peixinho}
\affiliation{Laboratoire PIMM, CNRS, Arts et M\'etiers Institute of Technology, Cnam, 151 boulevard de l’H\^{o}pital, Paris, France}


\begin{figure}
    \centering
    \includegraphics[width=1\textwidth]{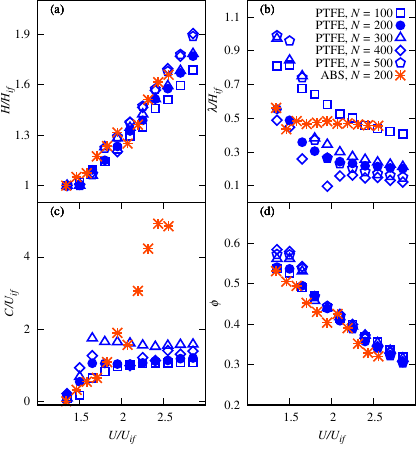}
    
    Fig. S1. Dimensionless dynamics of the particle beds. (a) Dimensionless time averaged bed height, $H/H_{if}$, (b) plug height, $\lambda/H_{if}$, (c) plug celerity, $C/U_{if}$ and (d) packing concentration, $\phi$, as a function of the upwards mean dimensionless water velocity, $U/U_{if}$. $H_{if}$ is the bed height and $U_{if}$ the water velocity at incipient conditions
\end{figure}

\begin{table*}[h]
    \centering
    \begin{tabular}{ccccccc}
    \hline
Sphere material & $d$ (mm) & $\rho$ (g/cm$^3$) & Asphericity (\%) & $R_a$ ($\mu$m) & $\mu_{p-p}$ & $\mu_{p-w}$\\
    \hline
ABS  & $5.91\pm0.01$ & $1.9\pm0.01$ & $0.06\pm0.02$ & $1.25\pm0.60$ & 0.126 $\pm$ 0.008 & 0.122 $\pm$ 0.008 \\
PTFE & $5.87\pm0.01$ & $2.33\pm0.01$ & $0.11\pm0.04$ & $0.60\pm0.21$ & 0.057 $\pm$ 0.007 & 0.090 $\pm$ 0.010 \\
    \hline
    \end{tabular} \\
    \vspace{5px}
    Tab. S1. Properties of the spheres: the diameter, $d$, the density, $\rho$, the asphericity, the average roughness, $R_a$, and the measured dry dynamic sliding friction between the particle and a wall of the same material, $\mu_{p-p}$ and between the particle material and the PMMA wall, $\mu_{p-w}$
\end{table*}

\begin{figure*}[ht]
    \centering
    \includegraphics[width=1.0\columnwidth]{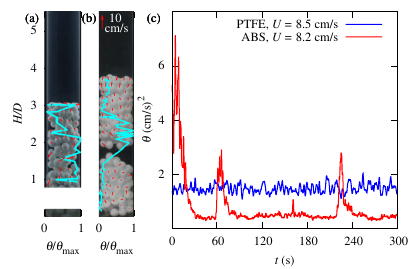}

    FIG. S2. Snapshots of particle velocities and normalized local granular temperature $\theta/\theta_{max}$ (a) for ABS with $N = 200$ and $U = 8.2$ cm/s and (b) for PTFE with $N = 200$ and $U = 8.5$ cm/s. The cyan lines represent the dimensionless local granular $\theta/\theta_{max}$ of the granular temperature, where $\theta_{max}$ is the local maximum temperature. The red arrows represent the particles velocity, a scale is displayed at the top of the snapshot. (c) Evolution of global granular temperature $\theta$ for PTFE and ABS
    \label{fig:crys-time}
\end{figure*}

\end{document}